\documentclass{appolb}
\usepackage{epsfig}
\usepackage{color}
\usepackage{cite}
\begin{document}
%\date{\today}
\pagestyle{plain}
% uncomment the following line to get equations numbered by (sec.num)
%\eqsec
\newcount\eLiNe\eLiNe=\inputlineno\advance\eLiNe by -1
\title {Ab initio analysis of all income society classes \\ in the European Union}
\author{M. Jagielski\footnote{e-mail: zagielski@interia.pl}, R. Kutner
\address{Institute of Experimental Physics \\ Faculty of Physics, University 
of Warsaw \\ 
Ho\.za 69, PL-00681 Warsaw, Poland}}
\maketitle

\begin{abstract}
%Herein, we applied statistical physics to study incomes of three society classes instead of the two classes 
%studied so far. 
We found a unified formula for description of the household incomes of all society classes, for instance, of those 
of the European Union in year 2007. This formula is a stationary solution of the threshold Fokker-Planck equation 
(derived from the threshold nonlinear Langevin one). The formula is more general than the well known that of 
Yakovenko et al. because it satisfactorily describes not only household incomes of low- and medium-income society 
classes but also the household incomes of the high-income society class.
\end{abstract}
\PACS{89.20.-a, 89.65 Gh\\
\begin{center}
Final draft submitted to Acta Physica Polonica A
\end{center} }

\section{Introduction}
In study of socio-economical systems, physics oriented approaches have widely been developed to explain different 
socio-economic processes \cite{Stan, Yako1, Yako2, RS, SR, RHCR,ME,ME1}. Those approaches aim at formulating well 
fitted unbiased indicators of social and economic phenomena. One of their key issues is the income of 
society analysis using methods of statistical physics, in particular, the stochastic dynamics considered as 
\emph{ab initio} level. The main goal of this economic issue is to unravel and describe mechanisms of societies' 
enrichment or impoverishment.

In the recent decade, a large number of studies were performed aiming at constructing of models, which 
(to some extend) would well  replicate the observed complementary cumulative distribution functions of individual 
incomes. Among them, the most significant seems to be the Clementi-Matteo-Gallegati-Kaniadakis approach \cite{CMGK}, 
the Generalized Lotka-Volterra Model \cite{SR, RS, RHCR}, the Boltzmann-Gibbs law \cite{ChCh, BChCh, DY,DY1}, and the 
Yakovenko et al. model \cite{Yako1, Yako2}. 
However, none of the above attempts to find an analytical description of the income structure solves the principal 
challenges, which concern:
\begin{itemize}
\item[(i)] the description of the annual household incomes of all society classes (including the third, i.e. 
the high-income society class) by a single unified formula based on the \emph{ab initio} level and
\item[(ii)] the problem regarding corresponding complete microscopic (microeconomic) mechanism responsible for 
the income structure and dynamics. 

In our considerations presented herein, we used Boltzmann-Gibbs law, weak Pareto law and Yakovenko et al. model 
to derive a uniform analytical formula describing all three society classes.
\end{itemize}

\section{Extended Yakovenko et al. model}
In accord with an effort outlined above, we compared the empirical data of the annual household incomes in the 
European Union (EU), including Norway and Iceland, with predictions of our theoretical approach proposed herein. This 
approach is directly inspired by the Yakovenko et al. model. By using the generalised assumptions we extended this 
model to solve our principal challenges (i) and (ii) indicated above.

We used data records from the Eurostat Survey on Income and Living Conditions (EU-SILC) \cite{EURO3}, by way of example 
for year 2007 \cite{EURO1} (containing around $200$ thousand empirical data points). However, these records contain 
only few data points concerning the high-income society class, i.e. the third region in the plot of the complementary 
cumulative probability distribution function vs. annual household income. To consider the high-income 
society class systematically, we additionally analysed  the effective income of  billionaires\footnote{The term `billionaire' used 
herein is equivalent (as in the US terminology) to the term `multimillionaire' used in the European terminology. Since 
we consider wealth and income of billionaires in euros, we recalculated US dollars to euros by using the mean exchange 
rate at the day of construction of the Forbes 'The World's Billionaires'.}$^{,}$\footnote{The billionaires who gained effective incomes are billionaires whose incomes are greater than zero.} in the EU by using the Forbes 'The World's 
Billionaires' rank \cite{Forbes}.    

We were able to consider incomes of  three society classes  thanks to  the following  procedure.
\begin{itemize}
\item[(i)] Firstly, we selected EU billionaires' wealth from the Forbes rank, for instance, for two successive years  
2006 and 2007.
\item[(ii)] Secondly, we calculated their incomes for year 2007. This calculation was possible because we assumed that 
billionaires' incomes  were proportional to the corresponding differences between their wealth for pair of successive 
years, here 2007 and 2006. Notably, we took into account only billionaires who gained effective incomes.
\item[(iii)] Subsequently, having calculated incomes for the high-income society class, we joined them with the EU-SILC dataset. By using so completed dataset, we then constructed the initial empirical complementary cumulative distribution function for year 2007. For that, we used the well known Weibull recipe \cite{WCH, Chow}. However, this direct approach shows a wide gap of incomes inside the high-income society class resulting in a horizontal line of the complementary cumulative distribution function. This gap separates the first segment belonging to the high-income society class, consisting of all data points taken from the EU-SILC dataset, from the second segment, consisting of remaining data points, which also belong to the high-income 
society class but are taken from the Forbes dataset. 
\item[(iv)] In the final step, we eliminated this gap by adopting the assumption that the empirical complementary cumulative distribution function (concerning the whole society) have no horizontal segments. That is, we assumed that statistics  of incomes is a continuous function of income (i.e. it has no disruption). Hence, we were forced to multiply the
billionaire incomes from Forbes dataset by the properly chosen common proportionality factor. This factor was equal to $1.0\times 10^{-2}$, as we assumed the requirement of full overlap of the first (above mentioned) segment by the second segment. This assumption leads to a unique solution (up to some negligible statistical error) for this proportionality factor. We found that this factor was only a slowly-varying function of time (or years). 
\end{itemize}

Hence, we received data record containing already a sufficient number of data points for all society classes, including 
the high-income society class. Although the Forbes empirical data only roughly estimate the wealth of billionaires, 
they quite well establish the billionaires' rank, thus sufficiently justifying our approach. This is because our 
purpose is to classify billionaires to concrete universality  class rather than finding their total incomes. 

The basic tool of our analysis is an empirical complementary cumulative distribution function being typical in this 
context. We calculated it according to the standard two-step procedure based on the well known Weibull formula 
\cite{WCH, Chow}. 
%For that, first, the income empirical data 
%were ordered according to their rank, i.e. from incomes of the richest households to those of the poorest. Next, in 
%accordance with the well known Weibull formula \cite{WCH, Chow}, we calculated the ratio $\frac{l}{n+1}$ where $l$ is 
%the position of the household in the rank and $n$ is the size of the empirical data 
%record. This ratio directly determines the required fraction of households of the income higher than that related to 
%a given household position $l$ in the rank. 
The complementary cumulative distribution function obtained that way is sufficiently stable and it does not reduce the 
size of the output compared to that of the original empirical data record.

Let $m$ be an influx of income per unit time to a given household. We treat $m$ as a variable obeying stochastic 
dynamics. Then, we can describe its time evolution by using the nonlinear Langevin stochastic dynamics equation 
\cite{Yako1, Yako2, Kamp}.  
%\begin{eqnarray}
%\frac{dm}{dt}=-A(m)+C(m)\, \eta(t).
%\label{rown1}
%\end{eqnarray}
%Here, $A(m)$ is a drift term and $\eta(t)$ is a temporal Gaussian noise, $N(0,1)$, where the coefficient $C(m)$ is 
%its $m$-dependent amplitude.
Hence, this Langevin equation is equivalent to the following Fokker-Planck  
equation for the probability distribution function (in the It\^o representation) \cite{Kamp}:
\begin{eqnarray}
\frac{\partial}{\partial t}P(m,t)&=&\frac{\partial}{\partial m}[A(m)P(m,t)]+
\frac{{\partial ^2}}{\partial m^2}\left[B(m)P(m,t)\right]. 
\label{rown2a}
\end{eqnarray}
%where the flux density of probability (in the It\^o representation \cite{Kamp}) is
%\begin{eqnarray}
%J(m,t)&=&-A(m)P(m,t)-\frac{{\partial}}{\partial m}\left[B(m)P(m,t)\right].
%\end{eqnarray}
Here, $A(m)$ is a drift coefficient and $B(m)=C^2(m)/2$, where the coefficient $C(m)$ is the $m$-dependent amplitude 
of a temporal white noise; they together play a fundamental role in the Langevin equation as a stochastic force. The 
quantity $P(m,t)$ is the temporal income distribution function. In general, coefficients $A(m)$ and $B(m)$ can be 
additionally determined by the first and second moment of the income change per unit time, respectively, only if these 
moments exist. Subsequently, the equilibrium solution of Eq. (\ref{rown2a}), $P_{\rm eq}$, takes the form:
\begin{eqnarray}
P_{\rm eq}(m)=\frac{const}{B(m)}\exp\left(-\int_{{m_{\rm init}}}^m\frac{A(m')}{B(m')}\, dm'\right)
\label{rown5}
\end{eqnarray}
where $m_{\rm init}$ is the lowest household income and $const$ is a normalisation factor. Indeed, this expression
is exploited in this work.
%Fortunately, both It\^o 
%and Stratonovitch representations \cite{Kamp} give almost the same equilibrium distribution function.

Following the Yakovenko et al. model \cite{Yako1, Yako2}, we can assume that changes of income of the low-income 
society class are independent of the previous income gained. This assumption is justified because the income of 
households belonging to this class mainly takes the form of wages and salaries. The stochastic process associated with 
the mechanism of this kind is called the additive stochastic process. In this case, coefficients $A(m)$ and 
$B(m)$ take, obviously, the form of positive constants
\begin{eqnarray}
A(m)=A_0,\quad B(m)=B_0.
\label{rown6}
\end{eqnarray}
This choice of coefficients leads to the Boltzmann-Gibbs law with exponential complementary cumulative 
distribution function \cite{Yako1, Yako2, ChCh, BChCh, DY, DY1}:
\begin{eqnarray}
\Pi(m) = \int_{m}^{\infty} P_{\rm eq}(m')\, dm' = \exp\left(-\frac{m-m_{\rm init}}{T}\right).
\label{rown7}
\end{eqnarray}
In Equation (\ref{rown7}), distribution function is characterised by a single parameter, i.e. an income temperature 
$T=B_0/A_0$, which can be interpreted in this case as an average income per household.

For the medium- and high-income society classes, we can assume (again following Yakovenko et al. \cite{Yako1, Yako2}) 
that changes of income are proportional to the income gained so far. This assumption is also justified because profits 
go to the medium- and high-income society classes mainly through investments and capital gains. This type of stochastic 
process is called the multiplicative stochastic process. Hence, coefficients $A(m)$ and $B(m)$ obey the proportionality 
principle of Gibrat 
\cite{Arma, Suto}:
\begin{eqnarray}
A(m)=a\, m,\quad B(m)=b\, m^2\Leftrightarrow C(m)=\sqrt{2\, b}\, m, 
\label{rown8}
\end{eqnarray}
where $a$ and $b$ are positive parameters. By using the equilibrium distribution function, Eq. (\ref{rown5}), we arrive in this 
case to the weak Pareto law with complementary cumulative distribution function \cite{Yako1, Yako2, RHCR}:
\begin{eqnarray}
\Pi(m) = \int_{m}^{\infty} P_{\rm eq}(m')\, dm' = \left(\frac{m}{m_{\rm s}}\right)^{-\alpha}.
\label{rown9}
\end{eqnarray}
Here, $m_{\rm s}$ is a scaling factor (depending on $a,\, b$, and $const$) while $\alpha=1+a/b$ is the Pareto exponent. The 
ratio of the $a$ to $b$ parameters can directly be determined from the empirical data expressed in the log-log plot 
(by using their slopes). 

As Yakovenko et al. have already found \cite{Yako1, Yako2}, the coexistence of additive and multiplicative stochastic 
processes is allowed. By assuming that these processes are uncorrelated, we get
\begin{eqnarray}
A(m)=A_0+am, \;
B(m)=B_0+b\, m^2=b\, (m^2_0+m^2),  
\label{rown10}
\end{eqnarray}
where $m^2_0=B_0/b$. This consideration leads (together with Eq. (\ref{rown5})) to a significant Yakovenko et al. model 
with the probability distribution function given by
\begin{eqnarray}
P_{\rm eq}(m) = const\, \frac{e^{-(m_0/T)\arctan(m/m_0)}}{[1+(m/m_0)^2]^{(\alpha +1)/2}},
\label{rown11}
\end{eqnarray} 
where parameters $\alpha$ and $T$ are defined above.

Based on the Yakovenko et al. Eq. (\ref{rown11}), the complementary cumulative distribution function can describe 
income of only low- and medium-income society classes. However, it does not capture that of the most intriguing 
high-income society class. 

The goal of our present work is to derive from Eq. (\ref{rown5}) such a 
distribution function, which would cover all three ranges of the empirical data records, i.e. low-, medium, and 
high-income classes of the society (including also two short intermediate regions between them). To do that, we 
have to provide function $A(m)$ in the threshold form: 
\begin{eqnarray}
A(m) &=& \left\{ \begin{array}{ll}
A^<(m)=A_0+a\, m & \textrm{if $m<m_1$} \\
{A^{\ge }(m)=A'_0}+a'\, m & \textrm{if $m\ge m_1$,}
\label{rown12}
\end{array} \right. \nonumber\\
B(m)&=&B_0+b\, m^2=b\, (m^2_0+m^2).
\end{eqnarray}

At the threshold $m_1$, there is a jump of the proportionality coefficient of the drift term. That is, this term 
abruptly changes from $a$ to $a'$ while the formalism of the income change remains the same 
for the whole society. This formalism is expressed by the threshold nonlinear Langevin equation where particular 
dynamics distinguishes the range of the high-income society class from those of the others.

The threshold parameter $m_1$ can be interpreted as 
a crossover income between the medium- and high-income society classes. Remarkably, both income crossovers $m_0$ and $m_1(\ge m_0)$ are exogenous parameters. They should be determined from the dependence of the empirical complementary cumulative distribution function on variable $m$ because both crossovers are sufficiently distinct.

Subsequently, by substituting Eq. (\ref{rown12}) into Eq. (\ref{rown5}), we finally get
\begin{eqnarray}
P_{\rm eq}(m) =\left\{ \begin{array}{ll}
c'\, \frac{\exp\left(-(m_0/T)\arctan(m/m_0)\right)}{[1+(m/m_0)^2]^{(\alpha +1)/2}}, & \textrm{if $m<m_1$} \\
c''\, \frac{\exp\left(-(m_0/T_1)\arctan(m/m_0)\right)}{[1+(m/m_0)^2]^{(\alpha_1 +1)/2}}, & \textrm{if $m\ge m_1$}
\end{array} \right. \label{rown19}
\end{eqnarray}
where $\alpha _1=1+a'/b$ and $T_1=B_0/A'_0$. Apparently, the number of free (effective) parameters driving the two-branch distribution function, Eq.(\ref{rown19}), is reduced because this function depends only on the ratio of the initial parameters defining the 
nonlinear Langevin dynamics. 

For $m_1\gg m_0$, the interpretation of the distribution function, Eq. (\ref{rown19}), is self-consistent, as required, 
because the two power-law regimes are well defined. Then, for instance for $m\gg m_0$, the second branch in 
Eq. (\ref{rown19}) becomes the power-law dependence driven by the Pareto exponent $\alpha _1$ different (in general) from 
$\alpha$.

Importantly, our analysis indicates that the existence of the third income region is already allowed by theory. We are 
following this indication below. 

\section{Results and discussion}
In principle, we are ready to compare the theoretical complementary cumulative distribution function based on our 
probability distribution function $P_{\rm eq}(m)$, given by Eq. (\ref{rown19}), with the empirical data for the whole 
income range. However, the analytical form of this theoretical complementary cumulative distribution function is 
unknown in the closed explicit form. Therefore, we calculate it numerically. The key technical question arises on how to fit this complicated 
theoretical function to the empirical data. The fitting procedure consists of three steps as, fortunately, all 
parameters are to be found (in principle) by using independent fitting routines, as follows.

In an initial step, we found approximated values of crossovers $m_0$ and $m_1$ directly from the 
plot of the empirical complementary cumulative distribution function (or empirical data). Thus, uncertainty of the 
$m_0$ and $m_1$ parameters did not exceed $10\%$, which was sufficiently accurate. Moreover, we took the exact value 
of the parameter $m_{\rm init}$ as the first point in the record of the empirical data. 

Secondly, we determined the temperature $T$ value by fitting the Boltzmann-Gibbs formula, Eq. (\ref{rown7}), to the 
corresponding empirical data in the range extending from $m_{\rm init}$ to $m_0$ (both found in the initial step).
Notably, we assumed that this formula could be characterised by a single temperature value since the society as a whole 
was considered to be in (partial) equilibrium during the whole fiscal year. That is, we further put 
$T_1=T\Leftrightarrow A_0'=A_0$.

At the third step, we determined exponents $\alpha$ and $\alpha_1$ by separately fitting the weak Pareto law 
to the empirical data for the medium- and high-income society classes, respectively.

\begin{figure}[ht]
\centering
\includegraphics[scale=0.45]{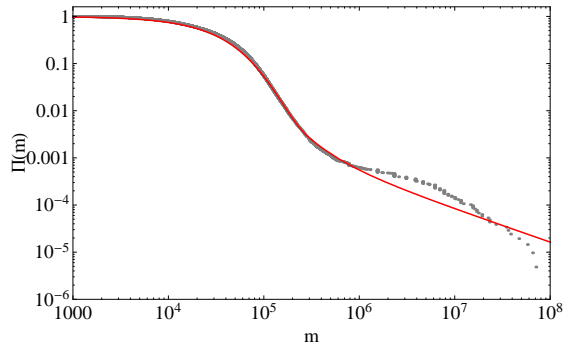}% Here is how to import EPS art
\caption{\label{fig1} Fit of the complementary cumulative distribution function, based on the extended Yakovenko 
et al. formula, Eq. (\ref{rown19}), (solid line) to the EU household income empirical data set (dots) for year 2007 
($T_1=T_2=T=37\times 10^3\pm 1\times 10^3$ EUR, $m_0=1.60\times 10^5\pm 0.16\times 10^5$ EUR, 
$m_1=3\times 10^5\pm 0.3\times 10^5$ EUR, $\alpha = 2.8643 \pm 0.0008$, and $\alpha_1 = 0.70 \pm 0.02$) 
\cite{EURO1, Forbes}.}
\end{figure}

Hence, we have already obtained all values required by the extended Yakovenko et al. formula, Eq. (\ref{rown19}). 
The corresponding plots of the empirical and theoretical complementary cumulative distribution functions in the log-log 
scale are compared in Fig. \ref{fig1}, for instance, for year 2007. 
Apparently, the predictions of the extended 
Yakovenko et al. formula, Eq. (\ref{rown19}), (solid curve in Fig. \ref{fig1}) well agree with the empirical data 
(dots in Fig. \ref{fig1}) for low- and medium-income society classes while agreement for the high-income society
class is satisfactory.

\section{Concluding remarks}

Herein, we proved that the household incomes of all society classes in the EU can be modelled by the nonlinear 
threshold Langevin dynamics with $m$-dependent drift and dispersion as \emph{ab initio} level. At the threshold $m_1$, 
there is a jump of the proportionality coefficient of the drift term. That is, this term abruptly changes from $a$ to $a'$, 
where $a'<a$ (as $\alpha _1<\alpha $). It means that the stochastic term in the Langevin equation is relatively more 
significant in this case (i.e. above threshold $m_1$) than the drift term. 

Furthermore, for the medium-income society class the Pareto exponent $\alpha>2$. This means that the variance of the 
Pareto distribution function exists and it is finite. However, for the high-income society class the variance of the 
Pareto distribution function is infinite, because $\alpha_1<1$. That is, assuming the variance as a measure of a risk, 
the economic activity of the high-income society class can be considered as more risky than activities of all other 
society classes, as expected \cite{Stan}.

The completed database, which we used (by properly joining the Forbes empirical database with that of 
EU-SILC), emphasises a significant role of the high-income society class. That is, only study of the income of 
all society classes enables adequate characterisation of the relative society wealth.

%The obtained results advance the knowledge on the subject matter \cite{ME,ME1}. We hope that these results will be useful for studies of static and dynamic properties of the household incomes not
%only for the EU as a whole but also for those of other continents and countries, if only sufficiently honest, large, 
%and complete (i.e. covering all society classes) empirical data are available. 

\section*{Acknowledgements}
We thank Victor M. Yakovenko and Tiziana Di Matteo for very stimulating comments and suggestions.

\end{document}